\documentclass[12pt]{article}
\let\Large=\large
\let\large=\normalsize

\newcommand{\be}[3]{\begin{equation}  \label{#1#2#3}}     
\newcommand{\ba}[3]{\begin{eqnarray} \label{#1#2#3}}
\newcommand{\bib}[3]{\bibitem{#1#2#3}}
\newcommand{\ee}{ \end{equation}}
\newcommand{\ea}{\end{eqnarray}}
\newcommand{\nn}{\nonumber}

\begin{document}

\thispagestyle{empty}
\rightline{HUB-EP-97/91}
\rightline{hep-th/9712142}

\vspace{1.5truecm}

\centerline{\bf \Large Entropy of bound states in the matrix description
            of M5-branes}
\vspace{2truecm}

\centerline{{\bf Klaus Behrndt}
\quad and  \quad
{\bf Ilka Brunner} \footnote{e-mail: 
behrndt@qft2.physik.hu-berlin.de;
brunner@qft3.physik.hu-berlin.de}
 }
\vspace{.5truecm}
\centerline{\em Humboldt University Berlin}
\centerline{\em Invalidenstrasse 110, 10115 Berlin, Germany}

\vspace{2.2truecm}


\vspace{.5truecm}

\begin{abstract}

\noindent
Like the M-theory itself also the worldvolume theory of the M5-brane
contains brane excitations, which can be extracted from the
supersymmetry algebra. Bound states (intersecting branes) of the
worldvolume can be translated into bound states of \mbox{11-d} SUGRA.
In this paper, we discuss the matrix description for these bound
states and their entropy (=degeneracy). In order to decouple the
worldvolume field theory from the bulk gravity we have especially to
assume that all charges are large, which gives a nice agreement with the
Bekenstein-Hawking entropy of black holes.
\end{abstract}

\bigskip \bigskip

\newpage

{\Large \bf 1. Introduction}

\medskip

The worldvolume theory of the M5 brane was the subject of many recent
publications \cite{140, 210, 220, 230}.  As introduction we will
briefly review some of the results of these papers.  The worldvolume
theory of an M5 brane is a (0,2) supersymmetric theory.  It has been
shown in \cite{140} that the most general (0,2) supersymmetry algebra
in 6 dimensions is of the following form:
\be600
\{Q_{\alpha}^I, Q_{\beta}^J\} = \Omega^{IJ} P_{[\alpha \beta]}
+ Y^{IJ}_{[\alpha \beta]} + Z_{(\alpha \beta)}^{(IJ)}
\ee
Here, $\alpha, \beta= 1, \dots 4$ are spinor indices of the 5+1
dimensional Lorentz group and  $I,J = 1, \dots 4$ are indices of the
R-symmetry group $Sp(2)$. The R-symmetry group has a geometrical
interpretation as the rotation group transversal to the M5 brane,
$Sp(2) \sim spin(5)$.  The charge $Y$ is a worldvolume 1-form 
($\sim (\gamma^m)_{\alpha\beta} Y^{IJ}_{m}$). It transforms as the
{\bf 5} of R-symmetry group. The other charge, $Z$, is a selfdual
3-form on the worldvolume ($\sim (\gamma^{mnp})_{\alpha\beta}
Z^{IJ}_{mnp}$) and transforms under the {\bf 10} of the R-symmetry
group. The existence of these charges suggest the possibilities of
p-branes, which are BPS states on the M5
worldvolume. Especially, in addition to strings on the M5 brane we
expect to have 3 branes living in the M5 brane worldvolume.  As was
pointed out in \cite{140} it is also possible to add a 5-form charge,
but this will not give additional degrees of freedom. A 5-form charge
leads to a worldvolume 5 brane filling the worldvolume of the M5 brane.
We can conclude from the algebra, that we have a certain
``degeneracy'': For each worldvolume direction we have 5 strings of
different R-charge and for three worldvolume directions we have 10
3branes of different R-charge.  

The properties of the algebra have a direct relation to the
classification of 1/4 BPS states in 11 dimensional M-theory.  We know
the ``intersection rules'' of the M5 branes with other M theory
branes, e.g. an M5 brane and an M2 brane can intersect over a string,
leading to a bound state at threshold. If the intersection region is
two dimensional, such that the M2 brane lies inside the M5 brane, only
a non-threshold bound state is possible.  The strings coming from
membranes intersecting M5 branes precisely correspond to the strings
expected from the algebra. We have already mentioned that the
R-symmetry group corresponds to the rotations in the transversal
space. Therefore, the R-charge of the strings leads to a space time
interpretation of the strings: The string states transform in the {\bf
5} representation of the R-symmetry group, which corresponds to the
fact that the membranes leading to the strings have one transversal
direction. Similarly, two M5 branes can intersect over a 3brane. 
It extends in three worldvolume directions and further two
transversal directions. There are 10 possibilities to pick two
transversal directions, corresponding to the fact that the charge $Z$
carries an index of the {\bf 10} representation of the R-symmetry
group.

All these states can also be identified by looking on the worldvolume
field theory, which has a selfdual 3-form field strength corresponding
to the string excitations and it has 5 scalars, which parameterize the
position of the brane in the 5 transversal directions. In order get
the 5-form field strength that couples to the 3brane we have 
to dualize one these scalars and, in addition, supersymmetry requires that
one has to turns on a further scalar.  Denoting these two scalars
by $U$ and $V$ the 3brane stretching along $x^1 , x^2 , x^3$
is described by the fields strength \cite{140}
\be601
G_{0123i} = \epsilon_{ij} \partial_j U = \pm \partial_i V \ .
\ee
where $i, j = 1,2$ and $U$ and $V$ are harmonic.  The gauge symmetry
implies that $U \simeq U + const.$ and $V \simeq V + const.$, where as
consequence of the Dirac quantisation the constants have to be integer
valued. Since these scalars parameterize the transverse position of
the brane, this means that a 3brane needs two compact transverse
directions and if these directions decompactify it becomes infinitely
heavy. This fits to the statement above, that the 3brane corresponds
to an intersecting 5brane that stretches also in $U$ and $V$, which are
called compact scalars of the 5brane worldvolume.

In our paper we will be especially interested in 4-dimensional black
holes coming from brane intersections involving a 5 brane.  As we have
seen, these brane intersections can also be interpreted as world
volume p-branes on the M5 brane.

The worldvolume theory of M5 branes also plays a role in Matrix theory
\cite{222}. In particular, there are ``little strings theories'' --
the iib strings discussed in \cite{240, 250} -- which also have
worldvolume threebranes and fivebranes. In the next section we will
discuss the (iib) strings as a Matrix model. We will in particular
discuss the Matrix picture of black holes involving 5 branes. In this
way, the pictures in space time and the Matrix model look very
similar. In the last section we compute the entropy of black holes
from the Matrix theory.

\bigskip \bigskip


{\Large \bf 2. Matrix description}


\medskip


In space time we want to consider (for example) the configuration of 3
M5 branes, where each pair of M5 branes intersects over a 3brane and
the common intersection of the three M5 branes is a string.  This
motivates us to look for a matrix description in terms of the M5 brane
with two compact scalars instead of the usual matrix description in
terms of the IIA D6 brane at strong coupling \cite{060}. On the
worldvolume of the M5 brane we have strings coming from membranes
wrapping one of the compact directions and we have 3branes from 5
branes wrapping both compact directions. So our base space
configuration looks very similar to the space time configuration.  

Let us first discuss the matrix description in terms of the M5 brane
with compact scalars. It was shown in \cite{160} that the matrix
description of M-theory on $T^4$ is given in terms of the strongly
coupled IIA D4 brane, which is better thought of as M5 brane.
In \cite{260} it was discussed that the light cone quantisation of
M-theory can be discussed in terms of an $\tilde{M}$ theory, which
is compactified on a space like circle. The limit we need to take in
$\tilde{M}$ is\footnote{We are grateful to A. Karch for numerous
discussions on this limit.}
\be070
R , l_p , L_i \rightarrow 0 \qquad  \mbox{but} \qquad \frac{R}{l_p^2} \  ,
 \  \frac{L_i}{l_p} \quad  {\rm fix}
\ee
As $R$ is taken to zero, M theory turns to IIA theory with
coupling $g_s^2=\frac{R^3}{l_p^3}$ and string tension 
$M_s^2 = \frac{R}{l_p^3}$. Because the transversal radii go to zero
we apply T-duality to come to a better description. In our
approach we apply T-duality 4 times giving us the IIA
D4 brane at strong coupling and the new IIA quantities are
\ba080
\Sigma_i &=& \frac{l_p^3}{R L_i} \\ \nn
M_s^2 &=& \frac{R}{l_p^3} \\ \nn
g_s^2 &=& \frac{l_p^9}{R V^2} \\ \nn
(M_P^{(10)})^8 &=& \frac{V^2 R^5}{l_p^{21}} \\ \nn
g_{YM}^2 &=& \frac{l_p^6}{R L_1 L_2 L_3 L_4},
\ea
where $M_P$ denotes the Planck mass and  $g_{YM} $ the Yang-Mills
coupling on the brane. This coupling is kept fixed in our limit, but
the string coupling goes to infinity. We therefore go to M-theory, which
leads to a new world-volume direction $\Sigma_5$, with 
$$
\Sigma_5 = \frac{g_s}{M_s} = g_{YM}^2 = \frac{l_p^6}{R L_1 L_2 L_3 L_4}
$$
The eleven dimensional Planck Mass is related to the ten dimensional
Planck mass by $ M_P^{(11)} = g_s^{-\frac{1}{12}} M_P^{(10)}$ and
using the relation (\ref{080}) we obtain for the new Planck
length $\tilde l_p \sim 1/M_P$
\be090
\tilde l_p^3 = \frac{l_p^9}{V R^2}
\ee
We want to consider this M-theory 5brane with two more compact 
directions. We didn't T-dualize them, so we have 
\ba100
U&=&L_5 \\ \nn
V&=&L_6
\ea
The limit (\ref{070}) becomes now
\ba110
\tilde l_p, U, V \rightarrow 0 \qquad {\rm but} \qquad
\frac{U}{\tilde l_p^3} \ ,\ \frac{V}{\tilde l_p^3} \quad {\rm fix}.
\ea
Note that these fixed quantities can be interpreted as tensions of strings
living on the M5 brane coming from membranes wrapping $U$
or $V$.
Taken together, the matrix model consists of
an M5 brane wrapping $\Sigma_1 \dots \Sigma_5$ with two more compact
directions $U$ and $V$ in the limit 
(\ref{110}). For convenience and later use we put together
the matrix-model/space time relations:
\ba120
\Sigma_i&=&\frac{l_p^3}{RL_i}, \quad i=1 \dots 4 \\ \nn
\Sigma_5&=&\frac{l_p^6}{R L_1 L_2 L_3 L_4} \\ \nn
U&=&L_5 \\ \nn
V&=&L_6 \\ \nn
\tilde l_p^3 &=& \frac{l_p^9}{L_1 L_2 L_3 L_4 R^2}
\ea
In the following we want to analyse the excitations of this model.
They should correspond to the energies of space time BPS states. 
In space time we expect 15 transversal wrapped membranes, 6 transversal
wrapped 5 branes and 6 momenta. For finite light like direction we
expect in addition 6 longitudinal membranes, 15 longitudinal 5 branes and
6 wrapped KK6 monopoles. Futhermore, we expect a KK6 brane
with NUT direction $R$ and a momentum mode of energy $1/R$.
We start with the transversal branes, which
correspond to bound states of the M5 to other branes in the matrix model.

First, we consider $n$ membranes stuck to the M5 brane. The energy of the
bound state is
\be130
\lim \sqrt{\left(\frac{N \, \Sigma_1 \Sigma_2 \Sigma_3 \Sigma_4 \Sigma_5}
{\tilde l_p^6} \right)^2 + \left(\frac{n \Sigma_i \Sigma_j}{ \tilde l_p^3 }
                        \right)^2}
-  
\frac{N \, \Sigma_1 \Sigma_2 \Sigma_3 \Sigma_4 \Sigma_5}
{\tilde l_p^6} = \frac{n^2 \Sigma_i^2 \Sigma_j^2}
{2N \,\Sigma_1 \Sigma_2 \Sigma_3 \Sigma_4 \Sigma_5}
\ee
Here, $\lim$ indicates the limit discussed in detail above
and $i,j \in {1 \dots 5}$. These bound states can also be interpreted
as fluxes of the field strength of the two form field living on the
world volume of the M5 brane. We now use (\ref{120}) to calculate the masses
of the corresponding space time BPS particles. For $i,j \neq 5$ we
obtain
\be140
M_{kl} = \frac{n L_k L_l}{l_p^3} \quad k,l \in {1, \dots 4} \quad k,l \neq i,j.
\ee
This gives us six transversal membranes. If $i$ (or $j$) equals $5$, we
obtain the kinetic energy of a particle with momentum in the
$L_j$ ($L_i$) direction:
\be150
E = \frac{n^2}{2 L_j^2} \, \frac{R}{N} = \frac{p_j^2}{2 p_{||}}.
\ee
This gives us four transversal momenta.

Next, we consider $n$ M5 branes. They can only form bound states of finite
energy with our ``fundamental'' M5, if they wrap one of the transversal
directions.  Taking $U$, the energy is:
\ba160
&& \lim \sqrt{\left(\frac{N \, \Sigma_1 \Sigma_2 \Sigma_3 \Sigma_4 \Sigma_5}
{\tilde l_p^6} \right)^2 + \left(\frac{n U \Sigma_i \Sigma_j
\Sigma_k \Sigma_l}{\tilde l_p^6} \right)^2}- 
\frac{N \, \Sigma_1 \Sigma_2 \Sigma_3 \Sigma_4 \Sigma_5}
{\tilde l_p^6} \\ \nn
&&~~~~~~~~~~~~~~~~~~~~~~~~~~~~~~~~~~~~~= 
\frac{n^2}{2N} \, \left(\frac{U}{\tilde l_p^3} \right)^2
  \, \frac{ \Sigma_1 \Sigma_2 \Sigma_3 \Sigma_4 \Sigma_5}{\Sigma_m^2}
\ea
In space time  for $m=1,2,3,4$ this leads to  membranes wrapping
$L_5$ and for $m=5$ we obtain a M5 brane wrapping $ L_1, \dots , L_5$.
Analogous results hold for a M5 brane wrapping $V$. In this
case we get four membranes wrapping $L_6$ and one direction in 
$L_1, \dots L_4$. In addition there is one M5 wrapping $L_1, \dots L_4, L_6$.

Further states can be obtained from Kaluza Klein monopoles. They have a
compact NUT-direction $\Sigma_i$ and stretch in six (other) directions.
Their tension is $\frac{\Sigma_i^2}{\tilde l_p^9}$ If the NUT direction
is one of the directions $1,\dots 5$ we can form bound states.
The energy of these bound states is
\be170
E= \frac{n^2}{2N} \, \left(\frac{U}{\tilde l_p^3} \right)^2 
 \, \left(\frac{V}{\tilde l_p^3} \right)^2 
\,  \Sigma_i^2 \Sigma_1 \Sigma_2 \Sigma_3 \Sigma_4 \Sigma_5.
\ee
Converting to space time quantities, we obtain for $i \neq 5$
\be180
M_i=\frac{n L_5 L_6 L_k L_m L_n}{l_p^6},
\ee
which are the masses of four transversal 5 branes. For $i=5$ we get
\be190
M_5=\frac{n L_5 L_6}{l_p^3}
\ee
which is the mass of $n$ membrane wrapping $L_5$ and  $L_6$.

Finally, we have to consider the momenta along $U$ and $V$.
They diverge, but give finite contributions together with the M5 brane.
In this way we find the two missing space time momenta.

In this way, we found all BPS states we expected in space time from the
matrix model.

Let us now come to objects whose energy is invariant under the limit
(\ref{110}) and it turns out that they can form bound states at 
threshold with the M5 brane.
In space time, they give branes wrapping the light cone direction and
in our case they are  M2 branes wrapping one of the compact directions and
M5 branes wrapping both compact directions. They correspond
to strings and threebranes living on the worldvolume of the M5 brane.
Furthermore, there are KK6 monopoles, whose NUT-direction is one
of the compact directions $U$ or $V$ and which wrap all
remaining  directions $\Sigma_i$. Especially, these branes stretch
in $\Sigma_1, \dots \Sigma_5$ and therefore look like 5 branes on
the M5 worldvolume.

Let us start with strings on the worldvolume. Taking $n$ strings 
coming from membranes wrapping $U$ and $\Sigma_i \, (i=1,\dots 4)$ 
have the energy
\be200
\frac{n U \Sigma_i}{\tilde l_p^3} = \frac{n R L_j L_k L_m L_5}{l_p^6}
\ee
and therefore is a description for $n$ space time longitudinal M5 brane.
Together with strings from membranes wrapping $V$ we obtained
8 longitudinal space time M5 branes.
For $i=5$ we get
\be210
\frac{n U \Sigma_5}{\tilde l_p^3} = \frac{n L_5 R}{l_p^3}
\ee
and an analogous formula for $V$. This gives us two longitudinal
membranes in space time.

Next, we consider 3branes on the worldvolume. Again,
3branes stretching along $\Sigma_5$ have a different space time
interpretation than 3branes only stretching in directions
$\Sigma_1, \dots, \Sigma_4$. Let us start with $i,j,k \neq 5$
Then the energy is
\be220
\frac{n U V \Sigma_i \Sigma_j \Sigma_k}{\tilde l_p^6}
=\frac{n L_5 L_6 L_i L_j L_k L_m^2 R }{l_p^9}
\ee
This means that in space time we have four KK monopoles with the NUT 
direction $L_m$
wrapping $L_i,L_j,L_k, L_5,L_6,R$. In the case that one of the
3brane directions, say $\Sigma_i$ is the $\Sigma_5$ direction, we
get:
\be230
\frac{n U V \Sigma_i \Sigma_j \Sigma_k}{\tilde l_p^6}
=\frac{n L_m L_n L_5 L_6 R}{l_p^6}
\ee
which are in space time 5 branes wrapping $L_5$ and $L_6$ and two
directions $L_m, L_n$ with $ m,n= 1\dots 4$ (altogether six M5 branes.).

Let us turn to 5 branes on the worldvolume coming from KK6 branes.
The energy for such a brane with NUT direction $L_{6,7}$ is
\be240
\frac{n \Sigma_{6/7}^2}{\tilde l_p^9} \, \Sigma_1 \Sigma_2 \Sigma_3
\Sigma_4 \Sigma_5 \Sigma_{7/6} = 
n \left( \frac{\Sigma_{6/7}}{\tilde l_p^3} \right)^2 \, 
\frac{\Sigma_{7/6}}{\tilde l_p^3} \, \Sigma_1 \Sigma_2 \Sigma_3 \Sigma_4
\Sigma_5 
\ee
giving wrapped KK6 with NUT direction 5 (6).

Finally, we have to take into account the momenta along the world
volume directions of the M5 brane. For $i=1, \dots,4$ they give 4
longitudinal membranes and for $i=5$ they lead to the longitudinal
M5 brane wrapping $R, L_1,\dots L_4$.

Altogether, we found 27 bound states corresponding to transversal branes
in space time and forming a {\bf 27} of the U-duality group
E$_{6(6)}$. This multiplet is the flux multiplet identified
in \cite{111}. The 27 bound states at threshold form the momentum multiplet
and correspond to longitudinal branes in space time. We can complete
the two times 27 states to 56 states if we take into account the
M5 brane itself, which corresponds to space time momentum along
the longitudinal R-direction and membranes wrapping $U$ and
$V$. These membranes have energies
\be250
\frac{n U V}{\tilde l_p^3} = \frac{n L_1 L_2 L_3 L_4 L_5 L_6 R^2}{l_p^9}
\ee
This is the energy of $n$ KK monopoles with NUT-direction R.
These states become light in our limit. In fact, Seiberg and Sethi 
\cite{333} have
argued that the M5 brane with two compact scalars does not decouple
from the bulk physics. This might be related to the appearance
of light states in the limit. However, if we consider infinite light
cone directions and infinite $N$, then we don't expect a matrix state
corresponding to a KK monopole with NUT direction $R$. In addition,
we have seen that the other states have finite mass in our limit
and in order to make sure, that they cannot leave the brane,
we will consider the limit of large charges where also their
masses become large. So in this case,
there might be hope that the worldvolume theory of the M5 including
3branes and two types of strings decouples from the bulk physics.

We are especially interested in configurations which are on the SUGRA
side intersections of three M5 branes, where each pair of M5 branes
have three world volume directions in common and all three M5's
intersect over a string. Along this direction, there is momentum.  We
take the common direction to be the light like direction R.  (It might
also be interesting to think about other directions.)  The momentum
along $R$ becomes in our matrix model the basic M5 stretching along
$\Sigma_1, \dots, \Sigma_5$, as can be seen from the relations
(\ref{120}). As we have seen, there are several possibilities to
produce longitudinal M5 branes from the matrix model.  Let us consider
the following space time configurations:
  \begin{center}
  \vspace{.2cm}
  \begin{tabular}{|c||c|c|c|c|c|c|c|}
  \hline
  &$x^1$&$x^2$&$x^3$&$x^4$ & $R$ & $x^5$&$x^6$\\
  \hline
  mom &&&&&o&&\\
  \hline
  M5 &x&x&x&x&x&&\\
  \hline
  M5 &x&x&&&x&x&x\\
  \hline
  M5 &&&x&x&x&x&x\\
  \hline
  \end{tabular}
  \vspace{.2cm}
\hfill
\vspace{.2cm}
\begin{tabular}{|c||c|c|c|c|c|c|c|}
  \hline
  &$x^1$&$x^2$&$x^3$&$x^4$ & $R$ & $x^5$&$x^6$\\
  \hline
  mom &&&&&o&&\\
  \hline
  M5 &x&x&x&&x&x&\\
  \hline
  M5 &x&x&&x&x&&x\\
  \hline
  M5 &&&x&x&x&x&x\\
  \hline
  \end{tabular}
  \vspace{.6cm}
  \end{center}
The matrix description of the first configuration is:
\begin{center}
 \vspace{.2cm}
  \begin{tabular}{|c||c|c|c|c|c|c|c|}
  \hline
  &$\Sigma_1$&$\Sigma_2$&$\Sigma_3$&$\Sigma_4$&$\Sigma_5$&$U$&$V$\\
  \hline
  M5 &x&x&x&x&x&&\\
  \hline
  mom &&&&&o&&\\
  \hline
  M5 &&&x&x&x&x&x\\
  \hline
  M5 &x&x&&&x&x&x\\
  \hline
  \end{tabular}
  \vspace{.2cm}
\hfill
 \vspace{.2cm}
  \begin{tabular}{|c||c|c|c|c|c|c|c|}
  \hline
  &$\Sigma_1$&$\Sigma_2$&$\Sigma_3$&$\Sigma_4$&$\Sigma_5$&$U$&$V$\\
  \hline
  M5 &x&x&x&x&x&&\\
  \hline
  M2 &&&&x&&x&\\
  \hline
  M2 &&&x&&&&x\\
  \hline
  M5 &x&x&&&x&x&x\\
  \hline
  \end{tabular}
  \vspace{.2cm}
  \end{center}  \vspace{.6cm}
{From} these tables one can directly read off how the branes and
momenta are mapped. Note, the $R$ direction in SUGRA is mapped on
$\Sigma_5$ and the momentum in SUGRA becomes our ``fundamental'' 5-brane
in the matrix description.

\bigskip \bigskip


{\Large \bf 3. The entropy of bound states}


\medskip

So far we described the matrix description of the $M5$-brane.  The
question is, whether it reproduces the same results then supergravity
or whether one can find new non-trivial statements.  Therefore we look
for quantities that can be compared. Obvious candidates are
(scattering) amplitudes and duality groups, but also the degeneracy of
bound states, i.e.\ the entropy, can be tested.  We will now discuss
the last option. Thus, we have first to find suitable bound states and
afterwards we have to discuss how we can determine their degeneracy.

Our matrix field theory is defined in a special limit (\ref{110}) of
the $M5$-brane worldvolume theory. So, we can take bound states of the
$M5$-brane theory and if they survive the decoupling limit they are
also bound states of the matrix description. There are two distinct
types of bound states, threshold and non-threshold. Threshold bound
states can naturally superposed, they have no binding energy, or in
other words, we can separate all constituents. They are much harder to
construct than the non-threshold bound states, which have a
non-vanishing binding energy.

A typical example for a non-threshold bound state of 11-d supergravity
is the configuration $2 \subset 5$, i.e.\ 2-brane lying inside a 5-brane
\cite{170} (for a collection of non-threshold $M$-theory bound states
see especially the second reference). This state and many others can
be obtained in SUGRA via $SL(2,R)$ transformation and/or boosts along
non-worldvolume directions of a single brane in 10 dimensions and
decompactify them to 11 dimensions.  This construction makes
clear, that they are less important for the discussion of  entropy,
which should be invariant under boosts and/or duality (when expressed
in terms of the ``right'' charges). 

For our purpose more interesting are threshold bound states. Like for
the non-threshold bound states, also these states can be constructed
first as bound state of the $M5$-brane and then perform the decoupling
limit (\ref{110}). If they survive this limit, we regard them as
matrix threshold bound states. Also here one has a construction
procedure in 10 dimensions. One can start with any known brane, makes
it non-extremal and perform a boost, but now along a worldvolume
direction. In contrast to the case above we can make here an infinite
boost while doing the extreme limit. Then after repeated $S$- and
$T$-duality transformations we get all known intersections. After
having a bound state of two branes we can repeat this procedure by a
boost along a common worldvolume direction. After decompactification
to 11 dimensions we get the $M$-theory threshold bound states. Of
course, alternatively one can apply the known intersection rules for
constructing intersecting brane. Note, these threshold bound states
break further supersymmetry, i.e.\ after compactification on a torus
they are not ``typical'' BPS states with 1/2 unbroken supersymmetry.

After we know of how to construct threshold bound states, we can turn
to the question of the degeneracy (=entropy) of these bound states. In
supergravity it should be given by the Bekenstein-Hawking
entropy of the black hole. In the matrix description, however, we have
a field theory without gravity and there is no obvious horizon. But
also in supergravity one does not need necessarily a horizon to obtain
the entropy. Instead the entropy can also be obtained from the minimum
of the BPS mass \cite{180}, a procedure that has been used extensively
for $N$=2 black holes, see e.g.\ \cite{190}. Therefore, the entropy
for a 4-charge configuration is given by
\be500
{\cal S} \sim \hat M_{min}^2 
\ee
with $\hat M_{min}$ is the minimum of a dimensionless mass, i.e.\
given by $\partial_{\phi^a} \hat M(\phi^a) = 0$ where $\phi^a$ denotes
the moduli.  In a thermodynamical approach \cite{130} this is
reflected in the first law of thermodynamics where an additional
moduli dependent term has been added.  But also in matrix theory this
approach has already been discussed in \cite{120}.

We will apply this procedure for the two configurations shown in the
table: $5\times 5 \times 5 +mom$ and the $5\times 2 \times 5 \times
2$. In the matrix description these states appear as $5 \times
3 \times 3 + mom.$ and $5\times 3 \times 1 \times 1$, where the ``5''
corresponds to a non-trivial 5-form charge. We should stress here,
that these are bound states in the matrix description, i.e.\ in the
so-called base space theory. We made the detour through threshold
bound states in $M$-theory only to show, that these configuration are
really bound states at threshold. Note, up to relabelling of coordinates
both configurations correspond to the {\em same} SUGRA
configuration. Since these states are at threshold we can simply add
up all mass contributions and obtain
\be510
\hat M_{ 5 \times 3 \times 3 + m}= \sum_i \hat M_i = V_i \, p^i + N 
\tilde l_p/\Sigma_5
\ee
where $V_1 = \Sigma_1 \Sigma_2 \Sigma_3 \Sigma_4 \Sigma_5/\tilde
l_p^5$, $V_2 = \Sigma_3 \Sigma_4 \Sigma_5 U V/\tilde l_p^5$, $V_3 =
\Sigma_1 \Sigma_2 \Sigma_5 U V/\tilde l_p^5$ are dimensionless
parameters and $p^i$ are the charges related to the branes and $N$ is
the momentum number. Note, that the momentum modes on the matrix
side correspond to one of the 5-brane charges on the SUGRA side
and the original $N$  became $p^1$ (see table).
The hat on the mass should indicate that we made
it dimensionless by $\tilde l_p$, because only the minimum of a
dimensionless quantity can be treated as entropy.  The mass formula
for the second configuration looks completely analog
\be520
\hat M_{5\times 3 \times 1 \times 1} = V_1 p^1 + V_2 p^2 + V_3 q^3 + 
V_4 q^4
\ee
where now $V_1 = \Sigma_1 \Sigma_2 \Sigma_3 \Sigma_4 \Sigma_5/\tilde
l_p^5$, $V_2 = \Sigma_1 \Sigma_2 \Sigma_5 U V/\tilde l_p^5$, $V_3 =
\Sigma_4 U/\tilde l_p^2$, $V_4 = \Sigma_3 V/\tilde l_p^2$ and $p^1$,
$p^2$ are the 5- and 3-form charges and the one-form charges are
$q^3$, $q^4$.

Before we can start with the extremization we have to discuss the
moduli of our model. Obviously, $U$ and $V$ are moduli, because they
appear already as scalar fields in the $M5$-brane worldvolume
theory. But in addition, because we consider the field theory 
on a compact space, also the (dimensionless) radii $\Sigma_i/ \tilde
l_p$ are moduli. Therefore, in the mass formulae the $V_i$ are the
moduli over which we have to minimise.  Immediately we see, that it
will yield a vanishing mass as minimum, which is certainly wrong.
This however is simply a consequence that we varied the complete
moduli space, but instead one has to keep fix the volume of the
moduli space - only internal deformations are moduli.
In our case this means that we have to keep fix
\be530
\Sigma_1 \Sigma_2 \Sigma_3 \Sigma_4 \Sigma_5 U V= \tilde l_p^7
\ee
i.e.\ in extremizing we do not change the the overall volume but allow
deformation of the different cycles. In SUGRA one has the same
constraint, namely that one fixes the Newton constant in 11 and in 4
or 5 dimensions or equivalently one fixes the asymptotic flat
Minkowski space. Only if one compactifies 10-d string theory, one
treats the volume of the compact space as moduli, which is related to
the dilaton moduli and which corresponds to a variation of $R_{11}$.
Using this constraint for (\ref{510}), i.e.\ $\Sigma_5 = \tilde l_p
V_1 V_2 V_3$ we find
\be540
\hat M_{ 5 \times 3 \times 3 + m}= V_i \, p^i + N/V_1 V_2 V_3
\ee
with $i = 1 .. 3$. This mass as function of $V_i$ has a non-trivial
minimum and with the ansatz $V_i = {c \over p^i}$ we find for
\be550
\partial_i \hat M_{ 5 \times 3 \times 3 + m}= p^i - N/V_1 V_2 V_3 V_i = 0
\ee
that $c^4 = p^1 p^2 p^3 p^4$ and therefore we get for the entropy
\be560
{\cal S} \sim \hat M^2_{min} = \sqrt{2 N p^1 p^2 p^3} \ .
\ee
Using the constraint (\ref{530}) for our second configuration, i.e.\ 
here $V_1 V_2 V_3 V_4 = 1$, yields also
\be570
\hat M_{ 5 \times 3 \times 1 \times 1}= V_1 \, p^1 +
V_2 \, p^1 +V_3 \, q^1 + q^2/V_1 V_2 V_3 \ .
\ee
Up to a charge redefinition ($p^3 \rightarrow q^1$, $N \rightarrow
q^2$) this is the same mass and thus it has also the same entropy
(\ref{560}).  Note, as discussed after eq.\ (\ref{250}) we can
trust the decoupling limit only for a large $R$ direction (large $N$
limit). In the matrix description this limit supresses transversale
membranes, which tend to become light otherwise.  The other branes
have still finite mass in our limit, but to make sure that they cannot
leave the brane, we consider the limit of large charges, which makes also
these branes heavy, i.e.\ $N , p^i \gg 1$. And really,
looking on the Bekenstein-Hawking entropy for 4-d black holes yields
exactly the same result, up to exchanging the momentum number with
one of the magnetic charges: $N \leftrightarrow p^1$.
Our matrix-configuration desribes in
SUGRA the configuration $5 \times 5 \times 5 + mom.$ and after
compactification to 4 dimensions one obtains for the entropy ${\cal S}
\sim \sqrt{N p^1 p^2 p^3}$, where $N$ is the electric charge
related to the momentum. Like in our matrix description, one can trust
this entropy only in the limit of large charges, namely: $N \gg p^i
\gg 1$. However the reasons are different here, they follow namely
from the requirement, that the low energy approximation still
holds. Note, the black holes appear as solution of the low energy
effective action.  To be concrete, the first relations controlls the
values of the scalar fields on the horizon, especially one keeps the
dilaton (higher genus corrections) under controll and the second
relation keeps the curvature on the horizon small, i.e.\ we can
neglect higher curvature correction ($\alpha'$ corrections).  We have
to keep in mind that the non-renormalization theorems concerns only
the lowest order in the effective action, e.g.\ higher curvature terms
are not protected. Alternatively, one can also argue, that near
the horizon the space time factorizes in $AdS_2 \times S_2$ and
the above limits ensure that in the string frame both curvatures
are small.

Finally, we have to discuss the counting of the state degeneracy. Only
this gives the justification to call the minimum of the mass
entropy. A more rigorous discussion of the degeneracy of our
configurations can be found in \cite{200}.  In our decoupling limit,
only string degrees of freedom can occur as dynamical modes (all 5-
and 3-branes masses are large).  These are just our momentum modes
travelling around the common intersection of the 3-branes.
Fortunately, for string states we know the degeneracy formula
\be033 
d(N) = e^{2 \pi\sqrt{c N/6}} 
\ee 
where $c$ is the central charge and in the case at
hand this is nothing as the effective dimension.
Now, the state counting can go in complete analogy to the D-brane
counting; keeping in mind that (i) the charge of the
brane are the number of branes that are on top of each other, (ii) the
effective dimension (where we can distribute the momentum modes) are
just the total number of layers, which is the product of charges and
finally (iii) taking into account bosonic and fermionic modes.
Doing all this, one finds an agreement with the entropy, that
we obtained as the minimum of the mass.

\bigskip

\bigskip

{\bf Acknowledgments}

We are grateful to A. Karch for many discussion and comments. 
The work is supported by the Deutsche Forschungsgemeinschaft (DFG).



\end{document}